\documentstyle[12pt]{article}

\newcommand{\newsection}{    % Numeration of eqs. is automatic
\setcounter{equation}{0}
\section}
\begin{document}
%%%%%%%%%%%%

\begin{titlepage}
{\bf August, 1992}\hfill	  {\bf LPTENS-92/23}\\
% hep-lat/9208023
\begin{center}

{\bf MIXED MODEL OF INDUCED QCD}

\vspace{1.5cm}

{\bf  A.A.~Migdal}\footnote{Permanent addres: Physics Department,  Princeton
University, Jadwin Hall,
Princeton, NJ 08544-1000. E-mail: migdal@acm.princeton.edu}

\vspace{1.0cm}

{\it Laboratoire de Physique Th\'eorique\footnote{Unit\'e propre du
CNRS, associ\'ee \`a l`Ecole Normale Sup\'erieure et \`a
l`Universit\'e de Paris Sud.}\\
de L'Ecole Normale Sup\'erieure, 24 rue Lhomond,\\
75231 Paris CEDEX 05, France}

\vspace{1.9cm}
\end{center}

\abstract{
The problems with the $Z_N$ symmetry breaking in the induced QCD are
analyzed. We compute the Wilson loops in the strong coupling phase,
but we do not find the $Z_N$ symmetry breaking, for arbitrary
potential.  We suggest to bypass this problem by adding to the model a
heavy fermion field in a fundamental representation of $ SU(N) $.
Remarkably, the model still can be solved exactly by the
Rieman-Hilbert method, for arbitrary number $N_f$ of flavors. At $ N_f
\ll N \rightarrow \infty $ there is a new regime, with two vacuum
densities.  The $Z_N$ symmetry breaking density satisfies the linear
integral equation, with the kernel, depending upon the old density.
The symmetry breaking requires certain eigenvalue condition, which
takes some extra parameter adjustment of the scalar potential.
}
\vfill
\end{titlepage}

%\tableofcontents

\newsection{Introduction}

\subsection{Critical phenomena in induced QCD}

The first few months of investigation of induced
QCD~\cite{KM92,Mig92a,Mig92b,GS92,Gr92,CAP92,KSW92,KhM92,KMSW92,Mig92c,Ko92,Ma92}
revealed some nice features of this new lattice gauge model, as well as some
problems. The functional integral of this model reads
\begin{equation}
  Z= \prod_{x} \int D \Phi_x \exp - N \mbox{ tr } \left[U(\Phi_x) \right]
\prod_{<xy>} I[\Phi_x,\Phi_y],
\end{equation}
with the Itzykson-Zuber integral
\begin{equation}
  I[\Phi_x,\Phi_y]\equiv\int D \Omega_{xy} \exp N \mbox{ tr } \left[
\Phi_x\Omega_{xy}\Phi_y\Omega_{xy}^{\dagger}\right],
\label{IZ}
\end{equation}
where the scalar field $ \Phi_x $ is the  $ N \otimes N $ hermitean
matrix and the gauge field $ \Omega_{xy} $ is the $ N \otimes N $ unitary
matrix. For the purposes of induction of QCD at large scales there is
nothing wrong with the $ U(1) $ subgroup of $ U(N) $, therefore we
leave the trace of $ \Phi $ finite, and $ \det \Omega \ne 1 $ to simplify
the equations.

The first  nice feature is the unique possibility to solve the model
analytically at $N = \infty $ by reducing it to the nonlinear integral
equation (master field equation) for the vacuum density $ \rho(\lambda) $ of
eigenvalues of the scalar field~\cite{Mig92a}, and the linear one for
the fluctuation $ \delta \rho_x(\lambda) $~\cite{Mig92b}.

The dimension $ D $ of the lattice enters as a
parameter in these equations. The critical phenomena  are quite rich:
there are infinite number of fixed points $ \rho_{\alpha}(\lambda) =
|\lambda|^{\alpha} $ , with different critical
indices  $ \alpha = 2n+1 \pm
\frac{1}{ \pi} \arccos{\frac{D}{3D-2}} $. Explicit massive solutions,
interpolating between pairs of the fixed points $ \rho_{\alpha}(\lambda)$ and
$\rho_{2k-\alpha}(\lambda) $ were found in the last
paper~\cite{Mig92c}. In particular, the masses of scalar excitations  scale as
\begin{equation}
  m_{phys}^2 \propto B^{\delta} \\;\;
 \delta = \frac{1}{2} + \frac{k-2m-1}{2(\alpha-k)},
\end{equation}
where $ B$ is the coeficient at the perturbation
\begin{equation}
  \rho(\lambda) = |\lambda|^{\alpha} + B |\lambda|^{2k-\alpha} \\;\;\alpha>k.
\end{equation}

At the critical point this coefficient vanishes, which is needed to
suppress the operator of the lower scaling dimension. One would expect
this $B $ to vanish as some linear superposition of deviations of the
parameters of the scalar field potential $ U(\Phi) = \frac{1}{2} m_0^2
\Phi^2 + \frac{1}{4!} \lambda_0 \Phi^4 + \dots $ from the critical
values. As it was discussed in~\cite{Mig92c}, the multicritical point of
the type $ \alpha,k,m $ is realized at $ \alpha > 2m+1 $ when the
terms $ O \left(\Phi^{2m} \right) $ in the scalar potential become
relevant in the wave equation.

The model, therefore, has a nontrivial thermodynamics with calculable
indexes in arbitrary dimension. The forbidden interval is $
\frac{1}{2} < D < 1 $, and the solutions at $ D>1 $ show no
pathologies, contrary to the string models. The string models can be
regarded~\cite{Ko92} as linear realizations of the gauge symmetry in the same
lattice gauge model, with the unitary measure replaced by the Gaussian
one, like in the Weingarten model. This leads to the singularity at $
D=1 $ with the forbidden interval $ D>1 $.

The difference between the unitary and the Gaussian measure is known
to produce an extra structure at a random surface in the strong
coupling expansion. In the weak coupling expansion, the usual
perturbative QCD is hopefully recovered in the vicinity of the unit
element, which  arises as the classical solution for the unitary
measure, but has no special role in case of the Gaussian one.

In a way, this field theory of the field $ \rho_x(\lambda) $, with
extra continuum coordinate $ \lambda $ represents the theory of
extended objects, with infinite internal motion in continuum limit,
when the support of eigenvalues extends to the whole real axis of $
\lambda $. This infinite internal motion is the only hope to recover
the perturbative QCD with the space independent master field. As it was
mentioned in~\cite{KM92}, the spectral integrals over $ \lambda$ in
the strong coupling expansion reproduce the lattice Feynman graphs of
the large $N$ theory, much in the same way, as it took place in reduced
Eguchi-Kawai models~\cite{Mig83}. The correspondence between the
master field of our model and quenched and twisted reduced models was
discussed in the recent paper~\cite{Ma92}.

\subsection{$ Z_N $ problem and loop averages}

These were nice features. The problems arise when one tries to
introduce quarks. Within the large $ N $ expansion the quarks act as
infinitesimal perturbation, which do not change the vacuum of the
theory. All the physics of the conventional $ \frac{1}{ N } $
expansion is described by the Wilson loops $ W(C) $ in fundamental
representation, averaged over QCD vacuum. The quark confinement
corresponds to the {\it global} $ Z_N $ symmetry of this vacuum. The
dynamical realization of this symmetry in QCD is the (minimal) area law $ W(C
\rightarrow \infty ) \sim \exp (- \sigma A(C)) $ at large distances,
combined with asymptotic freedom $ W(C \rightarrow 0 ) \sim 1 $ at
small ones.

It is not clear, how to get this behavior of the Wilson loop in our
model.  As it was noted already in the first paper~\cite{KM92}, and
discussed at length in subsequent papers~\cite{KSW92,KhM92,KMSW92}, the
effective action for the gauge field after elimination of the heavy
scalars involves only absolute values of the Wilson loops. This leads
to the spurious {\it local} $ Z_N$ symmetry\footnote{Actually, we get
the local $ U(1) $ symmetry, but we could always lower it to $ Z_N$ by
fixing the determinant of $ \Omega $.} in this model, as well as
in any other model, built from the traces in the adjoint
representation of the gauge group. As a consequence, we get the local
confinement: the Wilson loop vanishes, unless the loop folds on
itself, so that the minimal area equals zero:
\begin{equation}
  W(C) = \delta_{0,A(C)}
\label{A0}
\end{equation}
Formally, this corresponds to infinite string tension $ \sigma = \infty $.

Recently, the loop equations and the Eguchi-Kawai reduction were
studied for our model in~\cite{Ma92}. For the quadratic potential $
U(\Phi) $, the solution of the loop equations was found, in which the
adjoint Wilson loop, as well as the fundamental one, obey the trivial
"zero area law" (\ref{A0}), up to $ O \left( \frac{1}{ N^2 } \right) $
corrections.  The path amplitude of scalar field,
\begin{equation}
  \left \langle\Phi|G(L)|\Phi\right \rangle\equiv \left \langle  \Phi_{x_0}
\prod_{k=0}^{L-1} \Omega_{x_k,x_{k+1}}\otimes
\Omega^{\dag}_{x_k,x_{k+1}} \Phi_{x_L} \right \rangle =
\Lambda^L \left \langle \mbox{ tr } \Phi^2 \right \rangle,
\end{equation}
where the matrix $ \Phi $ is treated as a vector  in adjoint
representation. It is implied, that the path $ \{x_0,x_1,\dots,x_L\}
$ has no backtracking steps, and should it have, those steps would not
count, as the matrices would cancel. So, this amplitude depends upon
the algebraic
length~\cite{Mig83}, $ L(\Gamma) $  of the path $ \Gamma $, (as opposed to the
usual lattice length $ |\Gamma| = \mbox{\# of steps} $). The explicit
formula for $ \left \langle \mbox{ tr } \Phi^2 \right \rangle $, (in
our normalization of $ m_0 $ ),
\begin{equation}
  \left \langle \mbox{ tr } \Phi^2 \right \rangle
=\frac{2 D- 1}{m_0^2(D-1)+D\sqrt{m_0^4+4(1-2D)}},
\end{equation}
coincides with the semi-circle solution, found previously~\cite{Gr92}
from the master field equation.

The solution for $ \Lambda $ reads
\begin{equation}
  \Lambda=\sqrt{\frac{1}{4} m_0^4+(1-2D)}+ \frac{1}{2}m_0^2
\end{equation}
These solutions, however, fail to produce any critical
phenomena, which are necessary to match perturbative QCD. Apparently,
it is impossible to induce QCD without self-interaction of the scalar
field (if it is possible at all!).

The usual propagator $ \left \langle\Phi|G(x,y)|\Phi\right \rangle$ could be
obtained from this one by
summing over all paths
\begin{equation}
 \left \langle\Phi| G(x,y)|\Phi\right \rangle= \sum_{\Gamma_{xy}}
m_0^{-2|\Gamma_{xy}|}
\left \langle \Phi| G\left(L(\Gamma_{xy}) \right) | \Phi \right
\rangle
\end{equation}
This sum is calculable, but the results were not presented.
Presumably, they would  agree with the solution of the corresponding
wave equation~\cite{Mig92b,Mig92c}, inverting the Gaussian part of effective
Lagrangean for $ \delta \rho_x(\lambda) $.

The same quantities could be found for arbitrary potential from the
master field equation, by noting, that the one-link integrals
factorize over the path, so that the general invariant propagator
\begin{eqnarray}
  && \left \langle\lambda|G(L)|\mu\right \rangle\equiv
\left \langle  \frac{1}{\lambda-\Phi_{x_0}}
\prod_{k=0}^{L-1} \Omega_{x_k,x_{k+1}}\otimes
\Omega^{\dag}_{x_k,x_{k+1}}  \frac{1}{\mu-\Phi_{x_L}}\right \rangle \\
\nonumber & \,&  =
\left \langle \frac{1}{\lambda-\Phi_{x_0}}
\prod_{k=0}^{L-1} \left \langle \Omega_{x_k,x_{k+1}}\otimes
\Omega^{\dag}_{x_k,x_{k+1}} \right \rangle_{\Omega }
\frac{1}{\mu-\Phi_{x_L}} \right \rangle_{\Phi}.
\end{eqnarray}
where $ \left \langle \right \rangle_{\Omega} $ denotes the one-link  averages
with the measure
(\ref{IZ}), at fixed scalar
field, and $ \left \langle \right \rangle_{\Phi}$ denotes the scalar field
average. As before, $
\Phi $ is treated as a vector in adjoint representation. We could
write the following identity,
\begin{equation}
 \left \langle \Omega_{xy}\otimes \Omega_{xy}^{\dag} \right
\rangle_{\Omega} \frac{1}{\lambda -\Phi_{y}} =
 I \left[\Phi_x,\Phi_y\right]^{-1} \frac{1}{ \lambda- \nabla_{\Phi_x}}
I \left[\Phi_x,\Phi_y\right]= G_{\lambda}(\Phi_x)
\end{equation}
where $ \nabla_{\Phi} \equiv  \frac{1}{ N} \frac{\partial}{\partial
\Phi} $,
\begin{equation}
 G_{\lambda}(\Phi_x)= \left(\lambda - \nabla_{\Phi_x} - F(\Phi_x) \right)^{-1},
\end{equation}
and
\begin{equation}
  F(\Phi) = \left[\frac{1}{N}\frac{\partial\ln I \left[\Phi_x,\Phi_y
\right]}{\partial \Phi_x}\right]_{\Phi_x=\Phi_y=\Phi} =
\frac{U'(\Phi)}{2D} - \frac{1}{D} \, \wp \int \, d \nu
\frac{\rho(\nu)}{\nu- \Phi}.
\end{equation}
The details could be found in previous papers, or below, where we
repeat the same computations for the mixed model.
The function $G_{\lambda}(\Phi) $  was computed at $ N = \infty $
in~\cite{Mig92a}
\begin{equation}
  G_{\lambda}(\mu) =  \frac{1}{ \lambda - R(\mu)} \Re  \exp \left(
\int \frac{d\nu}{\pi(\nu - \mu - \imath 0 )} \arctan \frac{\pi
\rho(\nu)}{\lambda-R(\nu) } \right),
\end{equation}
with
\begin{equation}
  R(\nu) = \frac{U'(\nu)}{2D} + \frac{D-1}{D} \wp \int \, d\mu
\frac{\rho(\mu)}{\nu-\mu}
\end{equation}
Let us represent this function in terms of dispersion integral
\begin{equation}
  G_{\lambda}(\Phi) = \wp \int \, d\mu \frac{\left
\langle\mu|\Lambda|\lambda \right \rangle }{\mu-\Phi}
\end{equation}
with the kernel $ \left \langle\mu|\Lambda|\lambda\right \rangle$ given by
inverse dispersion
relation
\begin{equation}
  \left \langle\nu|\Lambda|\lambda\right \rangle= \wp \int \,
\frac{d\mu}{\pi^2}
\frac{G_{\lambda}(\mu)}{\nu-\mu},
\end{equation}
then we see, that the one-link average acts as the operator $\Lambda$
\begin{equation}
  \left \langle \Omega_{xy}\otimes \Omega_{xy}^{\dag} \right
\rangle_{\Omega} \frac{1}{\lambda -\Phi_{y}} = \int d\mu
\left \langle\mu|\Lambda|\lambda \right \rangle \frac{1}{\mu -\Phi_{x}}
\end{equation}
which yields the following result for the propagator
\begin{equation}
  \frac{\left \langle\nu|G(L)|\mu \right \rangle }{N} = \int d \lambda  \int d
\phi
\frac{\rho(\phi)}{(\nu-\phi)(\lambda-\phi) } \left
\langle\lambda|\Lambda^L|\mu \right \rangle
\end{equation}
In general, there is
always the trivial eigenvalue $ \Lambda_0 = 1 $, corresponding to the
unit matrix as an eigenfunction. The other eigenvalues must be less
than $ 1 $, otherwise there would be a phase transition.

In particular, according to~\cite{Mig92a}
\begin{equation}
 \left \langle \Omega_{xy}\otimes \Omega_{xy}^{\dag} \right
\rangle_{\Omega}\Phi_{y} = F(\Phi_x)
\end{equation}
For the quadratic potential, according to~\cite{Gr92}, $F(\Phi ) =
\Lambda \Phi $, so that there is a trivial linear eigenfunction $ \Phi
$ with the eigenvalue $ \Lambda $. This is in complete agreement with
the loop equations.

The same arguments lead to the following result for the adjoint
loop average
\begin{equation}
  W^{a}(C) = \delta_{0,L(C)} +\frac{1}{ N^2} \left[-1+
\left(1-\delta_{0,L(C)} \right) \mbox{ tr } \Lambda^{L(C)} \right]
\end{equation}
The leading term at $ N =
\infty $, as before, arises for the backtracking loop, and it
corresponds to the infinite string tension,
regardless the solution for the scalar field density. The $ \frac{1}{
N^2} $ correction describes the perimeter law for the pointlike
"mesons", propagating along the adjoint (double) path. The wave
equation, effectively summing over all these paths, was derived in the
previous papers~\cite{Mig92b,Mig92c}.

As for the fundamental Wilson loop, in the large $ N $ limit, we could
take the $ SO(2N+1) $ gauge group instead of the $ SU(N) $. The large
$N$ saddle point equations would be the same, up to $ O \left(
\frac{1}{N} \right) $ corrections. It is also known, that the loop
equations~\cite{Mig83} of the complete gauge theory, with the
Yang-Mills or Wilson terms, coincide with the same accuracy for the $
SU(N) $ and $ SO(2N+1) $ gauge groups.

On the other hand, there is no center in the $SO(2N+1) $ group, hence
the Wilson loop would not vanish. In virtue of the factorization
property~\cite{Mig83} $ \left \langle \mbox{tr}\,A \,\mbox{tr}\,B \right
\rangle=
\left \langle\mbox{tr}\,A\right \rangle\left \langle\mbox{tr}\,B \right
\rangle+ O(1) $ the fundamental Wilson loop
would be equal to the square root of the adjoint Wilson loop, in
agreement with suggestion of~\cite{KhM92}.  At $ N=\infty $,
therefore, both satisfy the zero area law, unless we invoke the subtle
mechanism of the spontaneous generation of the fundamental Wilson
loops in effective gauge action~\cite{KhM81,KhM92,Ma92}, which is not
clear how to do.

\subsection{Mixed model}

At this point it is worth recalling, that our model is somewhat
artificial.  There was no physical reason in the choice of the adjoint
representation of the matter field~\cite{KM92} to induce QCD. We did
so, simply because at that time this seemed to be the only model with
correct counting of powers of $N$, which could be solved by orthogonal
polynomial technique.  Later, when the model was actually solved, the
more powerful technique of the Rieman-Hilbert equations was found.
Now, when we are no longer limited to the Itzykson-Zuber integral, the time
came to extend the model.

We add to the model the heavy fermion constituent field, $ \Psi_x $,
with $ N_f \ll N \rightarrow \infty $ flavor components, and solve
this mixed model exactly.  The physical quarks $ q_x $ could be
introduced later, at larger spatial scales.  The $Z_N$ symmetry, as we
show, can be spontaneously broken in this solution, therefore the
physical, light quarks would be able to propagate in this vacuum.

By adjusting the parameters, we could interpolate between the local
confinement and free quark regimes, which gives us more chances to
induce QCD. At least, the most obvious objection is now eliminated.

The functional integral of the mixed model reads
\begin{eqnarray}
&&\prod_{x} \int D \Phi_x \exp - N \mbox{ tr }\left[ U(\Phi_x) \right]
 \int D \Psi_x \exp  N \mbox{ tr }\left[ M \Psi_x \bar{\Psi}_x \right]  \\
\nonumber & \,&
\prod_{<xy>} \int D \Omega_{xy} \exp N \mbox{ tr } \left[
\Phi_x\Omega_{xy}\Phi_y\Omega_{xy}^{\dag} +
\Omega_{xy} \Psi_y H_{xy}\bar{\Psi}_x +
 \Omega_{xy}^{\dag}\Psi_x H_{yx} \bar{\Psi}_y \right],
\end{eqnarray}
where  $ \Psi_x $($\bar{\Psi}_x$) is $ N \otimes N'
$($N'\otimes N$) matrix.
The second dimension $ N' =N_s\, N_f$ where $N_f$ is the number of
flavors, and $ N_s =  2^{\lfloor \frac{1}{2} D  \rfloor} $
is the number of spin components. The matrices $ H_{x, x+\mu} =
\frac{1}{2} \left(1+\gamma_{\mu} \right) $ act on the spin  components.

The physical quark confinement in this case would correspond to the
situation, where the masses of the $\bar{\Psi}q,\bar{\Psi}\Psi $ mesons would
stay in the lattice cutoff range in the local limit, when the masses
of physical $\bar{q}q $ mesons go to zero in lattice units.  In terms
of the Wilson loops, there would always be the perimeter law, with
large decrement, due to $ \bar{\Psi} $ following the quark along the
loop, but hopefully there would also be the term with area law, due to
the induced gluon planar graphs filling the loop of the light quark.

\newsection{Collective fields and classical dynamics at large $N$}

\subsection{Two densities}

The solution of the mixed model starts with the observation, that
one-link integrals
\begin{equation}
  I =\int d\Omega \exp N \mbox{ tr }\left[ \Phi_1 \Omega \Phi_2 \Omega^{\dag} +
\Omega
\Psi_2 H_{12} \bar{\Psi}_1 + \Omega^{\dag} \Psi_1 H_{21}\bar{\Psi}_2  \right],
\end{equation}
depend, in virtue of the gauge invariance, only upon the densities
$ \rho_1,\rho_2, \sigma_1,\sigma_2 $, where
\begin{equation}
  \rho_x(\lambda) = \frac{1}{N } \mbox{ tr } \delta(\lambda-\Phi_x),
\label{RO}
\end{equation}
\begin{equation}
  \left \langle i \alpha|\sigma_x(\lambda)|j \beta \right \rangle  =
  \frac{1}{N} \mbox{ tr } \delta(\lambda-\Phi_x) \Psi_x^{i,\alpha}
\bar{\Psi}_x^{j,\beta}.
\label{SI}
\end{equation}
with the spin indexes $ \alpha, \beta $ and flavor indexes $ i,j $
fixed. We shall treat $ \sigma $ as $ N'\otimes N' $ matrix.

This allows one to completely eliminate $ \Psi $ from the problem, by
the local change of variables\footnote{This simple observation,
applied to the old induced model, with quarks in place of $ \Psi $,
immediately rules out the
hypothesis, that it could induce QCD without the spontaneous breaking
of $Z_N$ symmetry.}. Let us introduce the $ N'\otimes N' $ matrix Lagrange
multiplier $ \epsilon(\lambda) $ for the constraint (\ref{SI}),
then we have to compute the integral
\begin{equation}
  \int D \Psi \int D \sigma \int D \epsilon \exp \left[ \mbox{ tr}'
\left( \Psi \epsilon(\Phi) \bar{\Psi}  \right)   - N\, \int d \lambda
\mbox{ tr}' \epsilon(\lambda) \sigma(\lambda)  \right],
\end{equation}
where $ \mbox{ tr}' $ corresponds to the spin and flavor trace, and the $
\epsilon $ integration goes along imaginary axis. The notation $
\epsilon(\Phi) $ is used for the matrix-valued function, in practice
this is used in the basis where $ \Phi $ is diagonal, where it means
the diagonal matrix of $ \epsilon(\lambda_i) $, where $ \lambda_i $
are the eigenvalues of $ \Phi $.

Integrating over $ \Psi $ first, we find
\begin{equation}
  \int D \Psi \exp \left[ \mbox{ tr}'
\left(\Psi \epsilon(\Phi)  \bar{\Psi} \right) \right] \propto
\prod_{i=1}^{N}  \det' \epsilon(\lambda_i)  =
\exp \left[  N \, \int d \lambda \rho(\lambda) \mbox{ tr}' \ln
\epsilon(\lambda)  \right].
\end{equation}
This yield the extra local term in effective action for the $
\epsilon, \sigma $ variables,
\begin{equation}
  \delta S_{eff}(\sigma,\epsilon,\rho) = N\, \int d \lambda \mbox{ tr}'
 \left( \epsilon(\lambda) \sigma(\lambda) - \rho(\lambda) \ln
\epsilon(\lambda)  \right).
\end{equation}
The local field $
\epsilon(\lambda) $ can be eliminated from equations of motion
\begin{equation}
  \epsilon(\lambda) = \rho(\lambda)\left(\sigma(\lambda)\right)^{-1},
\end{equation}
which yields
\begin{equation}
  \delta S_{eff}(\sigma,\rho) = \mbox{const } + N\, \int d \lambda
\rho(\lambda) \left(- N' \ln \rho(\lambda) +  \mbox{ tr}' \ln \sigma(\lambda)
\right).
\end{equation}

The extra term in effective action for the $ \rho $ field
was computed in~\cite{Mig92b}. It starts from $ O(N) $ terms in the
large $ N $ limit, which can be neglected in the leading order under
consideration, provided $ 1 \ll N_f $.
Therefore, the above term in effective action serves both
densities.

\subsection{Classical equations}

We assume, that $ 1 \ll N_f \ll N $, so that the flavor corrections are
important, but still the classical equations for the $ N'\otimes N' $
matrix $ \sigma $ could be applied.
Our objective now is to derive the set of equations for the one-link
integral, and solve them together with these classical equations. For
the space independent master fields $ \rho, \sigma $ the classical
equations read
\begin{equation}
 \frac{2D}{N^2}  \frac{d}{d \lambda} \frac{\delta \ln I}{\delta
\rho(\lambda)} + 2 \Re  V'(\lambda) =  U'(\lambda) + \frac{1}{N} \left(
- N'\,\frac{\rho'(\lambda)}{\rho(\lambda)} + \mbox{ tr}'
\frac{\sigma'(\lambda)}{\sigma(\lambda)}\right),
\label{EQRO}
\end{equation}
\begin{equation}
 \frac{2D}{ N^2 }\frac{\delta \ln I}{\delta \sigma(\lambda)} =
- M + \frac{1}{N}
\rho(\lambda)\left(\sigma(\lambda)\right)^{-1}.
\label{EQSI}
\end{equation}

The first classical equation follows from variation of the total
action with respect to $ \psi(\lambda) $, where $ \delta \rho = \psi'(\lambda)
$, the extra
derivative being introduced to preserve the normalization condition $ \int d
\lambda \rho(\lambda) = 1 $. The boundary conditions are $ \psi(\pm
\infty ) =0 $. This is the same equation as the old one,
except for the last term, coming from effective action. The potential
\begin{equation}
  V'(z) = \int d \nu \frac{\rho(\nu)}{z-\nu}
\end{equation}
and the values at the real axis are understood as limits from the
upper half plane
\begin{equation}
   V'(\lambda+\imath 0) = -\imath \pi \rho(\lambda) + \wp \int \, d \nu
\frac{\rho(\nu)}{\lambda -\nu}
\end{equation}

The second classical equation is new. It follows from the variation of
the total action in $ \sigma(\lambda) $, with spinor and flavor indexes implied
everywhere. The mass term and the scalar density are  proportional to
the unit matrix, and the second term in general case involves the
inverse of  $ \sigma $ matrix. We assume, that the vacuum densities $
\rho_x(\lambda)$ and $ \sigma_x(\lambda) $ are spatially homogeneous
and the second one is proportional to the unit matrix,
in virtue of the symmetry of the model.

\subsection{Schwinger-Dyson identities}

Apart from these classical equations, which are valid only at $ N
\rightarrow \infty $, there are some identities,
which are satisfied by the one-link integral. The simplest  is the
one, used in~\cite{Mig92a} to solve the scalar model,
\begin{equation}
 \frac{1}{ I}  \mbox{ tr }
\left(\lambda-\nabla_{\Phi_1} \right)^{-1} I= \mbox{ tr }
\left(\lambda - \Phi_2 \right)^{-1},
\end{equation}
where
\begin{equation}
   \left(\nabla_{\Phi}\right)_{ij} =  \frac{1}{ N }
\frac{\partial}{\partial \Phi_{ji}}.
\end{equation}
One may rewrite this identity as follows
\begin{equation}
   \mbox{ tr }\left(\lambda-\nabla_{\Phi_1}-F(\Phi_1) \right)^{-1}
=\mbox{ tr }  \left(\lambda - \Phi_2 \right)^{-1},
\label{ROTR}
\end{equation}
where
\begin{equation}
  F(\Phi_1) = \nabla_{\Phi_1} \ln I,
\end{equation}
is the matrix-valued function. In the same way, as
in~\cite{Mig92a,Mig92b}, we find
\begin{equation}
 2D F(\lambda) =  \frac{2 D}{N^2} \frac{d}{d \lambda} \frac{\delta \ln
I}{\delta \rho_1(\lambda)}= U'(\lambda)-2 \Re
V'(\lambda)  + \frac{1}{N} \left(
- N'\,\frac{\rho'(\lambda)}{\rho(\lambda)} + \mbox{ tr}'
\frac{\sigma'(\lambda)}{\sigma(\lambda)}\right)
\end{equation}

In order to compute another variation of $I$, we note, that in virtue
of unitarity of $ \Omega $,
\begin{equation}
  \mbox{ tr } \left(\lambda-\nabla_{\Phi_1}\right)^{-1}
\frac{\partial I}{\partial  \bar{\Psi}_1}   \bar{\Psi}_1 =
\mbox{ tr }
\left(\lambda-\Phi_2\right)^{-1} \Psi_2    \frac{\partial
I}{\partial \Psi_2},
\end{equation}
with sum over spins and flavors implied.
On the other hand, from definition of $ \sigma(\lambda) $ (for $\Psi
=\Psi_1,\Psi_2$)
\begin{equation}
  \frac{\partial I}{\partial \Psi} =  \frac{1}{ N} \bar{\Psi} \frac{\delta
I}{\delta \sigma(\Phi)} \\;\;
\frac{\partial I}{\partial \bar{\Psi}} =  \frac{1}{ N}  \frac{\delta
I}{\delta \sigma(\Phi)} \Psi,
\end{equation}
which yields, after moving $ I $ to the left, shifting $ \nabla
\rightarrow \nabla + F $, replacing the trace on the right by the
spectral integral,
\begin{equation}
  \frac{1}{ N} \mbox{ tr } \left(\lambda- \nabla_{\Phi}-F(\Phi) \right)^{-1}
\eta(\Phi)
=  \int d \mu  \rho(\mu) \frac{\eta(\mu) }{\lambda-\mu}
\label{EQJ }
\end{equation}
with
\begin{equation}
  \eta(\mu)=  \frac{1}{ N^2} \mbox{ tr}' \sigma(\mu)
\frac{\delta \ln I}{\delta \sigma(\mu) }
\end{equation}
{}From now on we are going to deal with classical solution $
 \sigma(\lambda) $ which has no indices in spin space.
Using the classical equation (\ref{EQSI})
\begin{equation}
  \eta(\mu)=\frac{N'}{2D}\left(-M \sigma(\mu) + \frac{1}{N}
\rho(\mu)  \right).
\end{equation}

We cannot directly represent the left of (\ref{EQJ }) in terms of the spectral
integral, as there are derivatives $ \nabla_{\Phi} $, which prevent us
from diagonalizing $ \Phi $. The problem is of the same kind, as in
the pure induced model, which we solved before. We have an identical
problem in the present model, in the scalar sector, described by
(\ref{ROTR}), so let us solve it first.

\subsection{First master field equation}

Let us introduce the matrix-valued function $ G^0_{\lambda}(\Phi) $ as
solution of the differential equation
\begin{equation}
 \nabla_{\Phi}G^0_{\lambda}(\Phi)=-1+\left(\lambda  -F(\Phi)
\right)G^0_{\lambda}(\Phi),
\end{equation}
then we have the following normalization condition from (\ref{ROTR})
\begin{equation}
  \mbox{ tr } \left( G^0_{\lambda}(\Phi) +  \frac{1}{\Phi-\lambda}
\right)=0
\label{TRG}
\end{equation}
which holds identically for any $ \lambda $. We choose $ \lambda $ to
belong to the support of the eigenvalues, and take the real solution
for $ G^0 $, corresponding to the principle value prescription.

In the large $ N $ limit, the derivative $ \nabla_{\Phi} $ acts as the
following integral~\cite{Mig92a}
\begin{equation}
  \nabla_{\Phi}G^0_{\lambda}(\Phi)= \int d \mu \rho(\mu)
\frac{G^0_{\lambda}(\mu)-G^0_{\lambda}(\Phi)}{\mu-\Phi}
\end{equation}
which reduces the problem to the Riemann-Hilbert integral equation
\begin{equation}
  1+\wp \int \, d \mu \rho(\mu)
\frac{G^0_{\lambda}(\mu)}{\mu-\nu}=(\lambda- R(\nu))G^0_{\lambda}(\nu),
\end{equation}
\begin{equation}
  R(\nu) = F(\nu) + \Re  V'(\nu)=\frac{U'(\nu)}{2D} + \frac{D-1}{D}
\Re  V'(\nu) + \frac{N_f\,N_s}{2N\,D} \left(-\frac{\rho'(\nu)}{\rho(\nu)}
+ \frac{\sigma'(\nu)}{\sigma(\nu)} \right).
\label{RLA}
\end{equation}

For the analytic function
\begin{equation}
  {\cal T}^0_{\lambda}(z) = 1+ \int \, d \mu \rho(\mu)
\frac{G^0_{\lambda}(\mu)}{\mu-z},
\end{equation}
which is defined in the upper halh plane and extended to the whole
plane by the  symmetry relation
\begin{equation}
  {\cal T}_{\lambda}(\bar{z}) = \bar{{\cal T}}_{\lambda}(z),
\end{equation}
this is the boundary problem
\begin{equation}
\frac{{\cal T}^0_{\lambda}(\nu+\imath 0)}{{\cal T}^0_{\lambda}(\nu-\imath 0)} =
\frac{\lambda-R(\nu)+ \imath \pi \rho(\nu)}{\lambda-R(\nu)-\imath \pi
\rho(\nu)}
\label{Bproblem}
\end{equation}
with the well known solution
\begin{equation}
{\cal T}^0_{\lambda}(z) = \exp \left( \int \frac{d \nu}{ \pi (\nu-
z )} \arctan \frac{\pi \rho(\nu)}{\lambda-R(\nu)}\right) .
\label{Solution}
\end{equation}
The solution for $ G^0_{\lambda} $ can be obtained from the real part
of the complex conjugate boundary values
\begin{equation}
  {\cal T}^0_{\lambda}(\nu \pm \imath 0 ) = \left( \lambda-R(\nu)\pm
\imath \pi \rho(\nu) \right) G^0_{\lambda}(\nu)\\;\;
G^0_{\lambda}(\nu)= \frac{\Re  {\cal T}^0_{\lambda}(\nu +\imath 0
)}{\lambda-R(\nu)}.
\end{equation}

As long as we are interested only in the equation for density, we do
not need this expression, but rather can use the asymptotic formula
\begin{equation}
  \lim_{z \rightarrow \infty} z \left(1-{\cal T}^0_{\lambda}(z)\right)=
\int \, d \mu \rho(\mu) G^0_{\lambda}(\mu)= \Re  V'(\lambda),
\end{equation}
where the last relation followed from (\ref{TRG}).
On the other hand, we can easily find the same quantity from the
explicit solution for $ {\cal T}^0_{\lambda}(z) $ which yields the
first master field equation
\begin{equation}
  \wp \int \, d \mu \left( \frac{\pi \rho(\mu)}{\mu-\lambda} + \arctan
\frac{\pi \rho(\mu)}{\lambda-R(\mu)} \right) =0,
\end{equation}
which differs from the old one by the last term in definition
(\ref{RLA}) of $ R(\lambda) $.

\subsection{The second master field equation}

Let us now derive the second master field equation. Repeating the same
steps, we find for the new function
\begin{equation}
  G_{\lambda}(\Phi)= \left(\lambda- \nabla_{\Phi}-F(\Phi)
\right)^{-1}\eta(\Phi)
\end{equation}
the following differential equation
\begin{equation}
 \nabla_{\Phi}G_{\lambda}(\Phi)=-\eta(\Phi)+\left(\lambda  -F(\Phi)
\right)G_{\lambda}(\Phi),
\end{equation}
and the  normalization condition
\begin{equation}
  \int d\mu \rho(\mu) G_{\lambda}(\mu) = \wp \int \, d\mu
\frac{\rho(\mu)\eta(\mu)}{\lambda-\mu}
\label{TRG1}
\end{equation}

Furthermore, at $ N = \infty $  we arrive at the integral equation
\begin{equation}
  \eta(\nu) +\wp \int \, d \mu \rho(\mu)
\frac{G_{\lambda}(\mu)}{\mu-\nu}=(\lambda- R(\nu))G_{\lambda}(\nu),
\end{equation}
with the same $ R(\nu) $ as before. The first term adds a new element
to the problem. Now, the analytic function
\begin{equation}
  {\cal T}_{\lambda}(z) = \imath \int \frac{d\mu}{\pi}
\frac{\eta(\mu)}{z-\mu}+  \int  d \mu \rho(\mu)
\frac{G_{\lambda}(\mu)}{\mu-z},
\end{equation}
in the upper half plane of $ z $, and it is continued to the lower
half plane by the symmetry, as before.
The boundary problem
\begin{equation}
  {\cal T}_{\lambda}(\nu \pm \imath 0 ) = \left( \lambda-R(\nu)\pm
\imath \pi \rho(\nu) \right) G_{\lambda}(\nu) \pm \imath \, \wp \int \,
\frac{d\mu}{\pi} \frac{\eta(\mu)}{\nu-\mu},
\end{equation}
is now an inhomogeneous one. It can be solved by substitution
\begin{equation}
  {\cal T}_{\lambda}(z) = {\cal T}^0_{\lambda}(z) {\cal J}_{\lambda}(z),
\end{equation}
which provides the following boundary problem for $ {\cal J}_{\lambda} $,
\begin{equation}
  \left( \lambda-R(\nu)\pm\imath \pi \rho(\nu) \right) \left( {\cal
J}_{\lambda}(\nu\pm \imath 0)G^0_{\lambda}(\nu) -G_{\lambda}(\nu) \right) =
\pm \imath  \wp \int \,
\frac{d\mu}{\pi} \frac{\eta(\mu)}{\nu-\mu}.
\end{equation}

This equation defines the imaginary part
\begin{equation}
  \Im  {\cal J}_{\lambda}(\nu + \imath 0)=  \frac{1}{ G^0_{\lambda}(\nu)} \,
\frac{(\lambda-R(\nu))}{(\lambda-R(\nu))^2 + \pi^2 \rho^2(\nu)}\,
 \wp \int \,
\frac{d\mu}{\pi} \frac{\eta(\mu)}{\nu-\mu}.
\end{equation}
so that we could restore $ {\cal J}_{\lambda} $ from dispersion relation
\begin{equation}
  {\cal J}_{\lambda}(z) = \int \frac{d\nu}{\pi} \frac{\Im  {\cal
J}_{\lambda}(\nu + \imath 0)}{\nu-z}.
\end{equation}
There are no subtraction terms here, as this function should decrease
as $ z^{-1} $ to satisfy the normalization condition. The coefficient
in front of $ z^{-1} $ can be found from above equations; this yields
the following relation
\begin{equation}
  \int \frac{d\nu}{\pi}\Im  {\cal J}_{\lambda}(\nu + \imath 0)=
 -\imath \int \frac{d\nu}{\pi}\eta(\nu) + \wp \int \, d\nu
\frac{\eta(\nu)}{\lambda-\nu}
\end{equation}

The real part of this relation provides us with the second master
field equation
\begin{equation}
  \wp \int d\nu\left(\frac{\rho(\nu)\eta(\nu)}{\lambda-\nu}
-\frac{1}{ G^0_{\lambda}(\nu)} \,
\frac{(\lambda-R(\nu))}{(\lambda-R(\nu))^2 + \pi^2 \rho^2(\nu)}\,
 \wp \int \,\frac{d\mu}{\pi^2} \frac{\eta(\mu)}{\nu-\mu}
\right) =0
\end{equation}

The imaginary part  yields
\begin{equation}
  0=\int d\nu\eta(\nu)
\end{equation}
or, substituting the explicit formulas,
\begin{equation}
   \int d\nu \sigma(\nu) = \frac{1}{N M}
\end{equation}
which is quite a surprise. The "heavy quark condensate" is trivialy
related to the bare mass. This is, actually, the lattice artifact,
nothing to do with critical phenomena. It is convenient to renormalize
$ \sigma(\lambda) \longrightarrow \frac{1}{N M} \sigma(\lambda) $,
$ \eta(\lambda) \longrightarrow \frac{N'}{2 D M} \eta(\lambda) $,
so that
\begin{equation}
  \eta(\lambda) = \rho(\lambda)-\sigma(\lambda) \\;\; \int d\nu
\rho(\nu) = \int d\nu \sigma(\nu) =1.
\end{equation}
The expression for $ R(\lambda) $ would not change, as it involves only
the logarithmic derivative of $ \sigma(\lambda) $, and the above
equation for $ \eta(\lambda) $ would not change, being linear.

Now, it is easy to see, that there is always the $ Z_N$ symmetric
solution
\begin{equation}
  \sigma(\lambda) = \rho(\lambda)
\end{equation}
of the master field equations, with the same scalar density $
\rho(\lambda) $ as before. The extra term in $ R(\lambda) $ exactly
vanishes in this case.

This is the old vacuum. One may readily check, that the average $
\left \langle \Omega \right \rangle_{\Omega} =0 $ in this case. In
general case, this average is linearly related to $ \eta(\lambda) $.
As for the adjoint averages, those would be the same, as before, as it
follows from above equations. The $ \frac{N_f}{N} $ corrections  drop from $
R(\lambda)
$, and $ \rho(\lambda) $ is the same, so the heavy fermions decouple
at infinite $ N $ and small $ \frac{N_f}{N} $.

However, with proper adjustment of parameters of the scalar potential,
the linear master field equation for $ \eta(\lambda) $ (with $\sigma =
\rho $ in $R $ ) could have a nonzero solution. This is the
spontaneous breaking of the $ Z_N $ symmetry in our model. After this
bifurcation point, the new vacuum, with two different densities would
be stable. Presumably, this one does induce QCD, but that remains to
be seen.

Let us stress once again, that this solution does not apply to the
Veneziano limit $ N_f \sim N $, as in this case the classical
equations for the matrix field $ \sigma $ are no longer valid.
However, the first correction in $ \frac{N_f}{N} $ which we found,
already breaks the $ Z_N $ symmetry of the vacuum.

\section{Acknowledgements}

I would like to thank Edouard Brezin, Volodja Kazakov, Ivan Kostov
and Samson Shatashvilli for interesting discussions.
This work was partially supported by the National Science Foundation under
contract PHYS-90-21984.

\newsection{Note added}

When this paper was finished, a new paper~\cite{KMSW92b} appeared,
where the trivial $ D=1 $ and Gaussian solutions were discussed at
great length. Also, the completeness of the Schwinger-Dyson equations,
leading to the master field equation, was questioned. Unfortunately,
this objection is based on a misunderstanding. The authors
of~\cite{KMSW92b} forgot about the gauge invariance, in virtue of
which the Itzykson-Zuber integral $ I[\Phi_x,\Phi_y] $ depend  on
$ \mbox{ tr } \Phi_x^n , \mbox{ tr } \Phi_y^m $, but cannot depend
upon $ \mbox{ tr } \Phi_x^n \Phi_y^m $, as they suggest. The subgroup
$ P_N $ of the gauge group independently permutes the eigenvalues $
\Phi_x^{(i)} $ and $ \Phi_y^{(j)} $, which eliminates the terms like $
\sum_j \Phi_x^{(j)} \Phi_y^{(j)} $.

So, the gauge invariance plays the role of the "mixed" Schwinger-Dyson
equations, they were worried about. With the gauge invariant Anzatz,
the Schwinger-Dyson equations unambiguously determine the Taylor
expansion of the logarithmic derivative $F$ of the Itzykson-Zuber
integral, as it was discussed in~\cite{Mig92a}. Actually, the
uniqueness of the reconstruction of $F$ is irrelevant, since we {\it
do not} solve the Schwinger-Dyson
identities for $F$, but rather substitute the classical equation for
$F$ into these equations, to obtain the master field equation for $
\rho $. All we need here, is the correct Schwinger-Dyson equation, the
completeness is not used. As for the other "lessons" from the
Gaussian and one-dimensional models, I doubt their relevance to the
problem of induction of QCD.

%\eop

\end{document}